# Evaluation of Computational Approaches of Short Weierstrass Elliptic Curves for Cryptography

*Kunal Abhishek[1], E. George Dharma Prakash Raj[2]*

[1]*Society for Electronic Transactions and Security, Chennai, India*
[2]*Bharathidasan University, Tiruchirappalli, India*
*E-mails*:    kunalabh@gmail.com    georgeprakashraj@yahoo.com

**Abstract**: *The survey presents the evolution of Short Weierstrass elliptic curves after their introduction in cryptography. Subsequently, this evolution resulted in the establishment of present elliptic curve computational standards. We discuss the chronology of attacks on Elliptic Curve Discrete Logarithm Problem (ECDLP) and investigate their countermeasures to highlight the evolved selection criteria of cryptographically safe elliptic curves. Further, two popular deterministic and random approaches for selection of Short Weierstrass elliptic curve for cryptography are evaluated from computational, security and trust perspectives and a trend in existent computational standards is demonstrated. Finally, standard and non-standard elliptic curves are analysed to add a new insight into their usability. There is no such survey conducted in past to the best of our knowledge.*

## 1. Introduction

Computation of elliptic curve requires extensive mathematical research to compute curve's parameters over large prime field for its use in cryptography [1]. There are several agencies like National Institute of Standards and Technology (NIST), Standards for Efficient Cryptography Group (SECG), Brainpool, etc., who have recommended standard elliptic curves over various prime field orders. However, it is important to note the rationale behind the approaches adopted for selection of elliptic curve parameters from computational, security and trust perspectives. The scope of this article is limited to the Short Weierstrass form of elliptic curves which are used for constructing most of the present cryptosystems such as Public Key Infrastructure (PKI) [2], Secure SHell (SSH), Transport Layer Security (TLS), IPSec, JSON Web Encryption (JWE) [3], etc.

The key contributions of this paper enlist:
 1. A comprehensive survey for evaluation of the computational approaches of cryptographically secure elliptic curves is presented.
 2. Evolution of Elliptic Curve Cryptography (ECC) with theoretical advancements in cryptographic mathematics and their significant impact on standardization of computational methods is presented.

3. Chronology of attacks on Elliptic Curve Discrete Logarithm Problem (ECDLP) and their countermeasures is presented.

4. Selection criteria of cryptographically secure elliptic curves are discussed.

5. A trend in computational approaches of elliptic curves in standards recommended by various agencies is demonstrated.

6. Standard and non-standard elliptic curves are compared from computational, trust and security perspectives to add a new insight into their usability.

Rest of the paper is organized as follows: Section 2 gives preliminaries on elliptic curves in Short Weierstrass form and ECDLP. Section 3 describes evolution of ECC with time and theoretical advancements in applied mathematics to establish present computational standards and selection criteria of elliptic curve. Section 4 focuses on evaluation of two popular approaches to compute cryptographically secure elliptic curves. Section 5 demonstrates the trend of approaches for computation of elliptic curve parameters adopted by various agencies in their proposed standards. Section 6 differentiates between standard and non-standard elliptic curves in various contexts. Finally, Section 7 concludes the paper with future directions.

## 2. Preliminaries

### 2.1. Elliptic curve in short weierstrass form

Let the finite field $\mathbb{F}_q$ has characteristic greater than 3. An elliptic curve $\mathbb{E}$ over $\mathbb{F}_q$ is the set of all solutions $(x, y)$ to an equation

(1) $$\mathbb{E}: y^2 = x^3 + ax + b,$$

where the coefficients $a, b \in \mathbb{F}_q$ and $4a^3 + 27b^2 \neq 0$, together with a special point $\infty$ called the point at infinity which serves as the identity element of $\mathbb{E}$ which is known to be an abelian group [4].

### 2.2. The elliptic curve discrete logarithm problem

**Definition 1 (ECDLP).** Given an elliptic curve $\mathbb{E}$ defined over a finite field $\mathbb{F}_q$, a point $P \in \mathbb{E}(\mathbb{F}_q)$ of order $n$, and a point $Q \in \langle P \rangle$, determine the integer $l \in [0, n-1]$ such that

(2) $$Q = lP.$$

The integer $l$ is called the discrete logarithm of $Q$ to the base $P$, denoted as $l = \log_P Q$ [5].

The definitions: Definition 2 [6], Definition 3 [7], Definition 4 [8] and Definition 5 [9] define supersingular curve, embedding degree, prime field anomalous curve and class number of elliptic curves respectively which need to be carefully considered for selection of elliptic curves with intractable ECDLP for cryptography.

**Definition 2 (Supersingular Elliptic Curves).** If $\#\mathbb{E}(\mathbb{F}_q) = q + 1 - t$ denote the order of elliptic curve then $\mathbb{E}(\mathbb{F}_q)$ is said to be supersingular if $p$ divides $t$ where $p$ be the characteristic of $\mathbb{F}_q$ and $t$ be the trace of $\mathbb{E}$.

$\mathbb{E}(\mathbb{F}_q)$ is supersingular provided the trace ($t$) of the curve, $t^2 = 0; q; 2q; 3q$ or $4q$ [6]. Supersingular elliptic curves are vulnerable to attack due to Menezes, Okamoto and Vanstone (MOV) which solves Discrete Logarithm Problem (DLP) of supersingular curves to the DLP in a finite field with sub-exponential complexity [6, 10].

**Definition 3 (Embedding Degree of Elliptic Curve).** If $\mathbb{E}(\mathbb{F}_q)$ be the elliptic curve over $\mathbb{F}_q$ then $\mathbb{E}$ is said to have embedding degree $k$, a smallest positive integer, such that $n \mid (q^k - 1)$ where $n$ be the base point order.

It is also observed that ECC standards do not allow elliptic curves with low embedding degrees.

**Definition 4 (Prime Field Anomalous Curves).** An elliptic curve $\mathbb{E}$ defined over a prime field $\mathbb{F}_p$ is said to be prime field anomalous if $\#\mathbb{E}(\mathbb{F}_p)=p$, i.e., the curve has trace 1.

Prime field anomalous curves are trace one curves for which the ECDLP can be solved in linear time [10]. The prime field anomalous attack does not extend to any other classes of elliptic curves but the one having trace one [8].

**Definition 5 (Class Number).** Let $h(N)$ denotes the class number of the order $N$ of elliptic curve $\mathbb{E}$. Then $h(N)$ is the minimum degree of a number field over which the elliptic curve $\mathbb{E}$ admits a faithful lift.

## 3. Evolution of elliptic curves for cryptography

Table 1. Evolution of Short Weierstrass elliptic curves for cryptography

| Year | Event | Impact on ECC Standardization |
|---|---|---|
| 1985 | Elliptic curves were proposed for use in cryptography | ECC were extensively studied to develop cryptosystems |
| 1987 | Efficient point counting algorithm on elliptic curves by Schoof, Elkies and Atkin called SEA Algorithm was developed [17-18] | Uses complexity $O(\ln^5 p)$ for point counting |
| 1992 | Elliptic Curve based Digital Signature Algorithm (ECDSA) was developed [19] | Considered as a mature signature scheme in NIST standard |
| 1993 | Reduction of ECDLP of supersingular elliptic curves having trace zero to logarithm in a finite field [6] | Became selection criteria for safe elliptic curve in all standards |
| 1994 | Proposal of Shor algorithm [20] generalizes to solve ECDLP Random Quantum Polynomial (RQP) time using quantum computers | Led to realization that elliptic curves will be unsafe once sufficient quantum capability is built. So, new computational standard required for quantum resistance |
| 1996 | It was proved that the condition $n \mid (q^k - 1)$ is sufficient to realize the MOV algorithm under mild condition. Further, it was proved that randomly generated curves have $k > \log^2 q$ [21] | Became selection criteria for safe elliptic curve in all standards |
| 1997 | Proposal of a linear algorithm to solve ECDLP of trace one [10, 22] | Became selection criteria for safe elliptic curve in all standards |
| 1999 | NIST recommendation of 15 elliptic curves [23] | Widely accepted standard later |
| 2000 | SECG recommendation of elliptic curves [24] | Widely accepted standard later |
| 2005 | Recommendation of Brainpool first set of elliptic curves for standardization [25] | International effort for elliptic curve standardization |
| 2010 | Brainpool revised their specifications and published Request for Comment (RFC) 5639 [26] | Standard established |
| 2014 | Review of existing elliptic curves generation mechanisms by Bernstein and Lange [27] who coined two terms: ECDLP security and ECC security. They observed that Short Weierstrass form of elliptic curves are dominant in both the software and hardware implementations | Two new terms: ECDLP security and ECC security became important verification criteria for curve selection with side channel attack resistance |
| 2014 | NUMS-curve (Nothing Upon My Sleeves) were proposed under IETF standard [28] | Curves with better performance proposed under IETF Standard |
| 2015 | NIST Call for next generation elliptic curves with new models and optimized parameters resistant to side channel analysis was placed [28] | NIST wanted to replace its standard elliptic curves |
| 2016 | NIST report [29] on Post Quantum Cryptography (PQC). Resistance of elliptic curve cryptosystems was looked for quantum computing | Isogenies of supersingular elliptic curves were discussed as resistant to PQC instead of ECDLP |
| 2017-2020 | Proposal of Quantum resources required to run Shor algorithm to solve ECDLP in polynomial time [30] | Roeteller et. al. suggested quantum resource estimates to break ECDLP |

Note: $N$=Order of elliptic curve, $q$=prime power, $k$=embedding degree.

Table 2. Chronology of attacks on ECDLP and their countermeasures

| Attack | Description type | Countermeasure type |
|---|---|---|
| Pohlig-Hellman, DLP attack | Private key can be recovered using Chinese Remainder Theorem [31] | $N$ must be a prime or near prime with small cofactor, $N \geq 2^{160}$ [5] |
| Pollard-rho, DLP attack | A parallelized Pollard-rho on $r$ processors can solve ECDLP in $\sqrt{(\pi n)}/\sqrt{(2r)}$ steps [5, 32] | $n \geq 2^{160}$ [13, 32] |
| Pollard's Lambda, DLP attack | Faster method than Pollard-rho when ECDLP lies in subinterval $[1, b]$ of $[1, n-1]$, where $b < 0.39n$ [13] | Private key should be selected uniformly at random within interval $[1, n-1]$ [30] |
| Index-Calculus, DLP attack | ECDLP can be solved using multiplicative group $\mathbb{F}_q^*$ of the finite field $\mathbb{F}_q$ [13] | Small prime fields should be avoided, i.e., $n \geq 2^{160}$ [13] |
| Exhaustive Search, DLP attack | Computes successive multiples of base point till public key is achieved | $n$ should be sufficiently large [8] |
| Shanks' Baby step Giant step, DLP attack | Fully exponential deterministic algorithm to determine $n$ on $\mathbb{E}(\mathbb{F}_q)$ which requires approximately $\sqrt{N}$ steps and around $\sqrt{N}$ storage | $n \geq 2^{160}$ [13] |
| Weil pairing and Tate pairing attacks, Pairing based attack | ECDLP of $\mathbb{E}(\mathbb{F}_q)$ can be reduced to ordinary DLP on extension field $\mathbb{F}^*_{q^k}$ for some $k \geq 1$ where the number field sieve algorithm can be used to solve ECDLP [4, 6]. MOV reduction attack [6] | $n \nmid (q^k - 1)\ \forall k \geq 20$ [7, 18] and $\forall k \geq (q-1)/100$ [5]<br><br>$p \nmid t$ and $t^2 \neq 0, 2q, 3q$ or $4q$ [6] (Non-supersingularity) |
| Multiple logarithm, DLP attack | Multiple instances of ECDLP for the same elliptic curve parameters | $n \geq 2^{160}$ |
| Prime field anomalous curve, Pairing based attack | Trace of $\mathbb{E}(\mathbb{F}_p)=1$, i.e., $\# \mathbb{E}(\mathbb{F}_p)=p$ [8, 12] | $N \neq p$ [5] |

Note: $q$=size of underlying field, $p$=prime characteristic, $n$=order of a point on $\mathbb{E}$, $N$=order of $\mathbb{E}$, $r$=number of processors, $k$=embedding degree, $t$=trace of curve.

Table 3. Elliptic curve parameters selection criteria

| Elliptic curve parameter | Criteria | Benefit(s) |
|---|---|---|
| Prime $p$ | 1. Crandall prime $2^\alpha - \gamma$ where $\gamma < 2^{10}$ [33, 34]<br>2. Montgomery-friendly prime $2^\alpha(2^\beta - \gamma) - 1$ where $\alpha, \beta, \gamma \geq 0$<br>3. $p \equiv 3 \mod 4$<br>4. Mersenne prime $p = 2^k - 1$<br>5. $p$= random value<br>6. Length of $p \geq 221$ bits [27] | 1. For best possible performance by limiting carry propagation during multiply-reduce and $\gamma$ is small [34]<br>2. Accelerates Montgomery arithmetic [33]<br>3. Such primes can compute modular square root in constant time countering constant time attack using Side channels [33]. The point compression method allows representing one point $(x, y)$ of $\mathbb{E}$ only its abscissa $x$ and one bit discriminating between the two possible values $\pm y$. However, recovering $y$ requires computing a square root in $\mathbb{F}_p$. This is easier when $p \equiv 3 \mod 4$ since in this case, $c^{(p+1)/2}$ is a square root of $c$ if $c$ is a square [9]<br>4. Mersenne primes are special primes of unique form which enables fast arithmetic [33]<br>Minimizes time for modular multiplication [35]<br>5. No pre-studied value or special structure vulnerable to cryptanalysis<br>6. To counter brute-force attack |
| Coefficient $a$ | 1. $a = -3$<br>2. $a$= random value | 1. For efficiency reasons. Practically all curves have low-degree isogenies to curves with $a = -3$, so this choice does not affect security. P1363 allows $y^2 = x^3 + ax + b$ without the requirement $a = -3$ [9]<br>2. No pre-studied value or special structure |
| Coefficient $b$ | 1. Should not be square in $\mathbb{F}_p$ [9]<br>2. $b$=random value | 1. To avoid compressed representations of elliptic curve points as $(0, 0)$ and $(0, x)$ would be identical as $x = \sqrt{b}$ with least significant bit as 0 [26]<br>2. No pre-studied value or special structure |
| Elliptic curve order $N$ | 1. $N$ should be prime [13, 18]<br>2. $N$ should be composite | 1. Prime order curve selected to resist Pohlig-Hellman and Pollard's Rho attacks [5, 9]. Small subgroup attacks are avoided [9, 13]<br>2. Prime group order curves do not have points with $y=0$ [36]. Special points of the form $(x, 0)$ exist if the curve has an even order [9] |
| Base point order $n$ | $n$ should be prime to avoid Weil and Tete pairing attacks [5, 9] | $n \geq 2^{160}$ and $n \nmid (q^k - 1)$ where $k$ is the embedding degree of elliptic curve |
| Cofactor $h$ | Preferably 1 | For optimal bit security, $h=1$ though $1 \leq h \leq 4$ for performance gain [5, 9, 36] |
| Base point $G_{x, y}$ | Randomly chosen base point [4] | Prime order of base point gives maximum elliptic curve group size |

Elliptic curves have been extensively studied and reviewed for cryptography soon after the proposals of Neal Koblitz and Victor Miller during 1985-1987. ECC has evolved with time and theoretical advancements in cryptographic mathematics, which subsequently has significant impacts on evolution of elliptic curve computational standards, which is discussed in Table 1. Moreover, elliptic curves are expected to be resistant to cryptographic attacks that can be ensured through the implementation of appropriate countermeasures. Table 2 [8] briefly depicts such countermeasures for important discrete logarithm (DLP) based attacks and pairing based attacks which resulted in the evolution of cryptographically safe elliptic curve selection criteria. Table 3 shows important selection criteria for elliptic curve parameters and their benefits to select elliptic curves with desired properties.

## 4. Evaluation of computational approaches

Elliptic curves need to qualify certain mathematical validations in order to certify that the elliptic curve has the claimed order, resists all known attacks on ECDLP and base point order has also the claimed order [5]. There are usually two approaches either of which can be used to compute an elliptic curve over prime field: first, the deterministic approach and second, the random approach. However, in both – the deterministic and random approaches, following conditions are critical for the elliptic curve to meet cryptographic requirements [4, 5, 11]:

C1: Resistance to Pohlig-Hellman and Pollard's Rho attack, i.e., $n>2^L$ where $n$ is sufficiently large prime that divides order of the elliptic curve group $\#\mathbb{E}(\mathbb{F}_q)$. Here, $L \geq 160$, the length in bits.

C2: Resistance to Semaev-Smart-Satoh-Araki attack (Smart-ASS) [10, 12], i.e., $L \leq \lfloor \log_2 q \rfloor$ ensures $2^L \leq q$ or $\#\mathbb{E}(\mathbb{F}_q) \neq q$. It avoids the attack on prime field anomalous curves.

C3: $n>4\sqrt{q}$ guarantees that $\mathbb{E}(\mathbb{F}_q)$ has a unique subgroup of order $n$ as $\#\mathbb{E}(\mathbb{F}_q) \leq (\sqrt{q}+1)^2$ by Hasse's theorem [5, 13] and so, $n^2 \nmid \#\mathbb{E}(\mathbb{F}_q)$.

### 4.1. Evaluation of deterministic approach

In this section, we evaluate the deterministic approach of computation of elliptic curves with respect to computational method, computational complexity, security, trust and specific gains for cryptography.

### 4.1.1. Computational method

Complex Multiplication (CM) is a widely accepted deterministic computational approach for standardization of elliptic curves. The CM method proceeds with fixing the prime field order p first and then constructs an elliptic curve over the field $\mathbb{F}_p$ [11]. It gives a choice for selecting primes of special forms, accepts the order of the elliptic curve field p as input, and determines the CM discriminant D. The field order $p$ is selected such that it meets the conditions C1, C2 and C3. The CM method is efficient when the finite field size $p$ and the field order $\#\mathbb{E}(\mathbb{F}_q)=p+1-t$ are chosen such that CM-field of $\mathbb{E}$, i.e., $\mathbb{Q}\left(\sqrt{(t^2-4p)}\right)$ has small class number [4, 5]. A crucial step of CM method is to compute the roots of a special type of class field polynomials called the Hilbert and Weber polynomials [14]. These polynomials are uniquely determined by $D$. Equations (3) and (4) [15], and (5) [16] constitute the basis of computation of Short Weierstrass elliptic curves using CM method.

**Definition 6 (Twist).** Given $\mathbb{E}: y^2=x^3+ax+b$ with $a, b \in \mathbb{F}_p$ the twist of $\mathbb{E}$ by $c$ is the elliptic curve given by

(3) $$\mathbb{E}_c: y^2=x^3+ax+b,$$

where $c \in \mathbb{F}_p$.

**Theorem 1.** If the order of an elliptic curve is $\#\mathbb{E}(\mathbb{F}_p)=p+1-t$, then the order of its twist is given as

(4) $$\begin{cases} \mathbb{E}_c(\mathbb{F}_p^*) = (p+1-t) \text{ if } c \text{ is square in } \mathbb{F}_p, \\ (p+1+t) \text{ if } c \text{ is non-square in } \mathbb{F}_p. \end{cases}$$

**Theorem 2 (Atkin-Morain).** Let $p$ be an odd prime such that

(5) $$4p=t^2+Ds^2,$$

for some $t, s \in \mathbb{Z}$. Then, there is $\mathbb{E}(\mathbb{F}_p)$ such that $\#\mathbb{E}(\mathbb{F}_p)=p+1-t$ [16].

The CM method is called the Atkin-Morain method when the elliptic curve is derived over prime field [37]. Equation (5) observes that $D$ is the integer which can be determined from a given prime p called the CM discriminant of $p$. Algorithm 1 describes a general CM method [38] for constructing an elliptic curve over a given prime field.

**Algorithm 1.** Elliptic curve generation over prime field using CM approach
*Input:* Nil
*Output:* Elliptic curve over a prime field $\mathbb{E}(\mathbb{F}_p)$
**Step 1.** Choose elliptic curve field order $p$, a prime
**Step 2.** Find smallest CM discriminant $D$ from equation (5) along with trace $t$
**Step 3.** Construct the orders of the two elliptic curve $\mathbb{E}(\mathbb{F}_q)=p+1\pm t$
**Step 4. if** one of the curve orders is a prime or nearly a prime
**Step 5.** Fix elliptic curve order
**Step 6. else** Repeat Step 1 to determine $D$ and $t$
**Step 7. end if**
**Step 8.** Construct the class polynomial $H_D(x)$ //Class polynomial is independent of p
**Step 9.** Find a root $j_0$ of $H_D(x)(\mod p)$ // $j_0$ is the $j$-invariant of the desired elliptic curve
**Step 10.** Set $k=j_0/(1728 - j_0)(\mod p)$ // such that $\mathbb{E}: y^2=x^3+3kx+2k$
**Step 11. if** $\#\mathbb{E} \neq p+1-t$
**Step 12.** Construct the twist $\mathbb{E}_c$ //using a randomly selected non-square $c \in \mathbb{F}_p$ following equations (3) and (4)
**Step 13**. **return** $\mathbb{E}_c$
**Step 14. else**
**Step 15. return** $\mathbb{E}$
**Step 16. end if**

4.1.2. Computational complexity

The bit complexity ($\beta$) of CM method depends on $b$ and $h$ where $b=$ length of field order $p$, $h=$ class number, $h_c=$ cross over class number for which the random approach and CM approach have the same runtime. When $h(D) < h_c(b)$ where $D$ is the CM discriminant, then CM method is faster than random approach [11]. CM method can generate a prime order elliptic curve in time $\tilde{O}((\log N)^4)$ [38].

4.1.3. Security

Deterministic approach is vulnerable to non-disclosed attacks. B e r n s t e i n et. al. [39] showed that standards can be sometimes purposely designed in such a way that it can be manipulated by the agency who recommended those standards. Also, sufficient information about the computational mechanisms of curve parameters has not been made publicly available [7]. It is always a concern for researchers that the ECDLP of deterministically computed elliptic curves can be solvable by using very efficient sub-exponential or polynomial time algorithm using non-guessable very high computing power unknown to outside world.

4.1.4. Trust

The elliptic curve parameters which are selected deterministically are sometimes distrusted due to lack of sufficient proofs of their computational mechanisms [40]. Moreover, trust in the curve parameters is doubtful due to possibility of intentional non-disclosed properties of the curve parameters. There are some serious statements of distrust expressed by many reputed scientists and researchers on NIST recommended elliptic curves which was generated through deterministic approach. Some of such statements of distrust are given as below:

- "I no longer trust the constants. I believe the National Security Agency (NSA) has manipulated them through their relationships with industry." – B r u c e S c h n e i e r (see [41]).
- "NIST should generate a new set of elliptic curves for use with ECDSA in FIPS 186... The set of high-quality curves should be described precisely in the standard, and should incorporate the latest knowledge about elliptic curves." – E d w a r d  F e l t e n (see [42, 43]).
- "NIST should ensure that there are no secret or undocumented components or constants in its cryptographic standards whose origin and effectiveness cannot be explained." – S t e v e  L i p n e r (see [42, 43]).
- "However, in practice the NSA has had the resources and expertise to dominate NIST, and NIST has rarely played a significant independent role." – K o b l i t z, K o b l i t z and M e n e z e s [7].
- "We don't know how $Q = [d]P$ was chosen, so we don't know if the algorithm designer [NIST] knows [the backdoor] $d$." – S h u m o w and F e r g u s o n (see [44]).
- "Consider now the possibility that one in a million of all curves have an exploitable structure that "they" know about, but we don't. Then "they" simply generate a million random seeds until they find one that generates one of "their" curves." – S c o t t [45].
- Many more.

### 4.1.5. Specific gains of deterministic approach

CM method adheres to "Performance over slightly sacrificed security" principle for computation of elliptic curves. Fast elliptic curve computation is possible in CM method due to elimination of the need for a point counting algorithm and fixing of certain parameters like prime $p$ with special structures [40]. CM method allows much faster arithmetic with elliptic curves as compared to random approach to achieve higher performance of elliptic curve cryptosystems [5]. It provides smaller, faster and easily implementable software code due to offline precalculations while adopting deterministic computational approach [46]. Prime order elliptic curves generated using CM method with $a= -3$ are backward compatible with implementation supporting most of the standardized elliptic curves [42]. CM method can only be adopted to construct ordinary elliptic curves with low embedded degree $k>6$ [7]. CM method is not efficient if there is no restriction on the class number of the elliptic curve [8]. This method is useful in deriving elliptic curves with small class numbers for which ECDLP is hard and gives the same security level as given by the elliptic curves which are generated randomly [5, 8].

### 4.2. Evaluation of random approach

Random approach allows obtaining elliptic curves, which are ordinary, and avoids any special form or structure. This approach uses 'early-abort strategy' to obtain desired elliptic curve [5]. A general observation is that elliptic curves generated using random approach have not been given preference for standardization. We evaluate random approach from computational method, computational complexity, security, trust and specific gains perspectives in this section.

### 4.2.1. Computational method

In random approach, the elliptic curve generation algorithm computes curve parameters keeping ECDLP security and procedural transparency in consideration. Algorithm 2 describes a general random approach as preferred in [3-6, 11, 17, 18, 27, 33, 38] to derive cryptographically safe elliptic curve over prime field.

> **Algorithm 2.** Elliptic curve generation over prime field using random approach
> *Input:* Randomness
> *Output:* Elliptic curve $\mathbb{E}(\mathbb{F}_p)$, base point $G_{x,y}$, curve order $N$
> **Step 1.** Select randomly a prime $p$ of desired size
> **Step 2.** Fix $K=GF(p)$      // Generate Field $K$ of order $p$
> **Step 3.** Choose randomly coefficient $a$
> **Step 4.** Choose randomly coefficient $b$
> **Step 5.** Generate $\mathbb{E}(K)$      // Elliptic curve over $\mathbb{F}_p$
> **Step 6.** if $4a^3+27b^2 \neq 0$   // Non-singularity check
> **Step 7.** else go to Step 3
> **Step 8.** end if
> **Step 9.** Compute order $N$ of $\mathbb{E}$
> **Step 10.** if $N$ is prime  // To resist Pohlig-Hellman attack
> **Step 11.** else go to Step 3
> **Step 12.** end if

**Step 13**. **if** $\mathbb{E}$ is supersingular   // To resist MOV attack
**Step 14. else** go to Step 3
**Step 15. end if**
**Step 16. if** $N \neq p$  // Non-anomalous check
**Step 17. else** go to Step 3
**Step 18. end if**
**Step 19.** Select randomly a base point $G_{x,y}$ on $\mathbb{E}$
**Step 20.** Compute base point order $n$    // $n \geq 160$ bits and $n > 4\sqrt{p}$
**Step 21. if** $n \neq N$   // Check for cofactor as 1
**Step 22. else** go to Step 19
**Step 23. end if**
**Step 24.** Compute Twist $\mathbb{E}_c$   // For twist security of elliptic curve
**Step 25.** Compute order $N'$ of $\mathbb{E}_c$
**Step 26. if** $\mathbb{E}_c$ is non-singular & $N'$ is prime & $\mathbb{E}_c$ is non-supersingular  // All criteria to be met for $\mathbb{E}_c$
**Step 27. else** go to Step 3
**Step 28. end if**
**Step 29. return** $\mathbb{E}(\mathbb{F}_p)$, $G_{x,y}$, $N$ // Return elliptic curve parameters

Here, the prime field $p$ is fixed and coefficients $a$ and $b$ are kept varying until a suitable elliptic curve $\mathbb{E}$ with prime order $N$ is obtained. Some validations to meet the cryptographic requirements C1, C2 and C3 are also kept. We observe that all the elliptic curve parameters such as $p$, $a$, $b$ and $G_{x,y}$ are randomly generated in order to avoid any special structure or known values whose choices are ambiguous.

### 4.2.2. Computational complexity

For random approach, the bit complexity ($\beta$) only depends on length of prime ($r_0$) and falls in the range $O(\log^{5+\epsilon} k_0 r_0)$ to $O(\log^7 k_0 r_0)$ where $\epsilon > 0$ and $k_0$ is the cofactor [11].

### 4.2.3. Security

Random approach does not allow any special structure of curve parameters in order to eliminate doubts on intentional non-disclosure of backdoors [5]. Elliptic curves, which are randomly computed, have no hidden goals that can be proved in determination of the curve parameters. This ensures that the elliptic curve parameters are trusted and not suspected to belong to a (not publicly known to be) vulnerable class. This approach is favourable when long-term security is desired with an ignorable sacrifice of efficiency [7]. Elliptic curves can be frequently changed for security reasons when computed randomly [40]. The only way to compromise elliptic curve security in such case is to solve ECDLP rather than just attacking particular classes of weak elliptic curves. Hence, random approach is specifically preferred to obtain elliptic curves for implementation in strategic or military grade cryptosystems.

### 4.2.4. Trust

Random approach ensures that no intentional construction with hidden weakness in the elliptic curve parameters is present in order to prevent future exploitation to recover user's private key [5]. The trust in derivation of the elliptic curve parameters is maintained due to the use of absolutely new values drawn randomly each time. Moreover, there are no patent issues with randomly selected new curve parameters. Random approach protects against attacks in special classes of elliptic curves, which may be vulnerable in future [5]. However, random values of elliptic curve parameters are always arguable by others for their emanation and random number generation, in case they are not explained adequately.

4.2.5. Specific gains of random approach

Random approach adheres to the principle of "security over performance" for computation of elliptic curve parameters. Computing order of the elliptic curve is a time-intensive task and hence, selecting elliptic curve using random approach is a slower process as compared to the deterministic approach where one starts with fixing the order of the elliptic curve. Point compression and decompression also require more computation in randomly generated elliptic curves [40]. Elliptic curves are computed with nearly the same probability to ensure that curves are not special in any sense when they are computed randomly [5, 11].

## 5. Approaches adopted by agencies for elliptic curve computation

Many agencies have recommended elliptic curves over various security levels for standardization. Table 4 depicts the popular standard elliptic curves in Short Weierstrass form with their computational approaches. Here, randomly generated elliptic curves means those elliptic curves whose parameters like field order p, field coefficients *a, b* and basepoint $G_{x,y}$ are randomly or pseudo-randomly (a secure hash function is used to generate curve parameters from random value given as input to the hash function to confirm that parameters are indeed computed pseudo randomly) generated or otherwise, they are considered to be obtained from the deterministic approach. Clearly, from Table 4, the trend demonstrates that the CM method, i.e., the deterministic approach is the preferred computational approach for standardization of elliptic curves.

Table 4. Computational approach adopted for Short Weierstrass elliptic curve computation

| Name of elliptic curve | Agency | Year | Security level in bits | Approach |
|---|---|---|---|---|
| NIST [23] | National Security Agency (NSA) | 2001 | 112, 128, 192, 256 | Deterministic |
| Brainpool [25, 26] | European Consortium of Companies and Government | 2005 | 128, 192, 256 | Pseudo-random |
| ANSSI FRP256v1 [39] | ANSSI | 2011 | 128 | Random |
| SECG [24] | Certicom | 2000 | 112, 128, 192, 256 | Deterministic |
| NUMS-Curves [28, 42] | Microsoft Research | 2014 | 128, 192, 256 | Deterministic |
| Russian Standardized Curves [47] GOST R 34.10-2001 GOST R 34.10-2012 GOST R 34.11-2012 | Russian National Cryptographic Standards | 2001, 2012 | 128, 256 | Deterministic |

## 6. Standard and non-standard elliptic curves

Elliptic curves are standardized to enable compatibility and interoperability across diverse applications. Moreover, non-standard elliptic curves are mostly used by strategic applications such as military applications or non-military but other critical infrastructure applications such as nuclear reactors' command and control systems etc. These applications do not really believe in Kerckhoffs's principle [48] of security, which says "A cryptographic system should be secure even if everything about the system, except the key, is public knowledge". Unlike Kerckhoff's principle, the strategic applications do believe that not only the keys but the algorithm too should also be kept private to protect critical information infrastructure better. In such cases, they compute elliptic curves preferably using random approach instead of deterministic approach. Table 5 compares between the standard and non-standard elliptic curves from computation, trust and security perspectives to help the readers about their usability concerns.

Table 5. Standard elliptic curves versus non-standard elliptic curves

| Standard elliptic curve | Non-standard elliptic curve |
|---|---|
| Prefers deterministic approach of computation to get performance benefits in elliptic curve arithmetic. This helps in standardization of elliptic curves by global acceptance | Prefers random approach of computation for long term security so that any special kind of curve is avoided which may lead to vulnerability to an unanticipated attack |
| Adheres to Kerckhoffs's principle of security and fixes elliptic curves for compatibility and interoperability among diverse applications across the globe | Adheres mostly to strategic principle of security which says that keys and algorithm both needs to be kept secret |
| Standard elliptic curves are subject to public exposure and often attract cryptanalysis as more people use it. Hence, there is always a high chance of collision with the secret key [49] | Negligible chance of collision with the secret key that's why random approach is preferred |
| Distrust comes with presence of special structures of the curve parameters | Trusted new values of curve parameters known to designer only. Prefers random approach to compute elliptic curve parameters |
| Standard elliptic curves are globally accepted and trusted | Not published and mostly not supported by the standards. Hence, trusted by their proposers or/and in closed group only |
| Compatible across applications and interoperable due to standardization | Not compatible. Applications need to be made interoperable explicitly |
| Better approach in case where elliptic curve needs to be computed over large prime fields | Better approach in case where elliptic curve needs to be transparently computed without any special structures known to others [50] |
| Curve parameters and compression techniques have patent issues | No patent issues |
| Already published and analysed thoroughly. Non deniable chances of hiding backdoors | Derivation procedure of curve parameters are known to the proposers only and hence, negligible chances of backdoors. High degree of trust observed by the proposers of non-standard elliptic curves |
| Standard elliptic curves are fixed to maintain compatibility among applications | Non-standard elliptic curves have edge over the standard ones as they can be replaced frequently for added security |
| More prone to get attacked by sophisticated advancements in mathematics and discoveries | In case of randomly selected curve parameters, curve is safe until sub-exponential algorithm is known to break it in particular [33] |

## 7. Conclusion and future directions

Short Weierstrass elliptic curves are widely used for cryptographic purposes. An evolution chart of events is presented which has significant impact on introducing elliptic curves for use in cryptography. We discuss about important attacks on ECDLP and their countermeasures, which became the basic selection criteria of elliptic curves for their consideration in cryptography. This paper also discuss rationale behind the selection criteria used to compute cryptographically suitable elliptic curve parameters. Two popular approaches, i.e., deterministic and random approaches to compute cryptographically secure Short Weierstrass elliptic curves and rationale behind them are evaluated in detail. A trend of approaches for computation of elliptic curve parameters for cryptographic purposes is also demonstrated which favours deterministic approach in standardization so far. We also differentiate between standard and non-standard elliptic curves with respect to their computational approaches, trust and security and bring out the desirable facts to choose either of them on need basis. Hence, it is inferred that this comprehensive evaluation and analysis of computational approaches of cryptographically safe elliptic curves will be helpful to those who wish to compute Short Weierstrass elliptic curves for design of cryptosystems with desired properties of the elliptic curves.

Standardization of elliptic curves, which are computed using random approach will be, preferred in future citing the trust requirements of strategic applications.

*Acknowledgements*: The authors would like to thank SETS for giving the opportunity to extensively work on elliptic curves and to write this article. Authors would also like to show their gratitude to Dr. Ananda Mohan P. V. and Dr. Reshmi T. R. for their valuable suggestions.

R e f e r e n c e s